\documentclass[reprint,aps,prl,superscriptaddress,nofootinbib,preprintnumbers]{revtex4-1}

\usepackage{comment}
\usepackage{graphics}
\usepackage{bm}
\usepackage{color}
\usepackage{amsmath}
\usepackage{amssymb}
\usepackage{mathrsfs}
\usepackage{graphicx}
\usepackage[caption=false]{subfig}

\begin{document}
\title{One-Electron Quantum Cyclotron as a Milli-eV Dark-Photon Detector}

\date{\today}

\author{Xing Fan}
\email{xingfan@g.harvard.edu}
\affiliation{Department of Physics, Harvard University, Cambridge, Massachusetts 02138, USA}
\affiliation{Center for Fundamental Physics, Department of Physics and Astronomy, Northwestern University, Evanston, Illinois 60208, USA}
\author{Gerald Gabrielse}
\email{gerald.gabrielse@northwestern.edu}
\affiliation{Center for Fundamental Physics, Department of Physics and Astronomy, Northwestern University, Evanston, Illinois 60208, USA}
\author{Peter W.~Graham} 
\email{pwgraham@stanford.edu}
\affiliation{Stanford Institute for Theoretical Physics, Department of Physics, Stanford University, Stanford, California
94305, USA}
\affiliation{Kavli Institute for Particle Astrophysics \& Cosmology, Department of Physics, Stanford University, Stanford, California 94305, USA}
\author{Roni Harnik}
\affiliation{Superconducting Quantum Materials and Systems Center (SQMS), Fermilab, Batavia, Illinois 60510, USA}
\affiliation{Theoretical Physics Division, Fermi National Accelerator Laboratory, Batavia, Illinois 60510, USA}
\author{Thomas G. Myers}
\affiliation{Center for Fundamental Physics, Department of Physics and Astronomy, Northwestern University, Evanston, Illinois 60208, USA}
\author{Harikrishnan Ramani} 
\email{hramani@stanford.edu}
\affiliation{Stanford Institute for Theoretical Physics, Department of Physics, Stanford University, Stanford, California 94305, USA}
\author{Benedict A. D. Sukra} 
\affiliation{Center for Fundamental Physics, Department of Physics and Astronomy, Northwestern University, Evanston, Illinois 60208, USA}
\author{Samuel S. Y. Wong} 
\affiliation{Stanford Institute for Theoretical Physics, Department of Physics, Stanford University, Stanford, California 94305, USA}
\author{Yawen Xiao} 
\affiliation{Stanford Institute for Theoretical Physics, Department of Physics, Stanford University, Stanford, California 94305, USA}

\begin{abstract}
We propose using trapped electrons as high-$Q$ resonators for detecting meV dark photon dark matter. When the rest energy of the dark photon matches the energy splitting of the two lowest cyclotron levels, the first excited state of the electron cyclotron will be resonantly excited.  A proof-of-principle measurement, carried out with one electron, demonstrates that the method is background free over a 7.4 day search.  It sets a limit on dark photon dark matter at 148 GHz (0.6~meV) that is around 75 times better than previous constraints. Dark photon dark matter in the 0.1--1 meV mass range (20--200~GHz) could likely be detected at a similar sensitivity in an apparatus designed for dark photon detection. 
\end{abstract}

\preprint{FERMILAB-PUB-22-482-SQMS-T
}

\maketitle

The particle nature of dark matter (DM) and its interactions with the standard model (SM) of particle physics remains a mystery, despite decades of experimental scrutiny~\cite{DMDiscovery1933,DMRotationCurve1980,DMGravitationalLensing2006,DMBulletCollision2000,DM_WMAPGalaxyCenter2007,Planck2018-1}. The mass of the DM is unknown and the possibility that it is made of ultralight bosons and can be described as a classical wave has received significant inquiry in recent years~\cite{FuzzyCDM2000,DarkPhotonMisalignmentMechanism2011,ATheoryOfDarkMatter2009,WISPyDarkMatter2012,ReviewScalarFieldDarkMatter,UltraLightScalarCosmologicalDarkMatter}. One such ultralight dark matter candidate is the dark photon (DP), a hypothetical spin-1 particle~\cite{DPTheory1986,Okun:1982xi} that is theoretically well motivated and possesses cosmological production mechanisms that can produce the observed DM abundance~\cite{Graham:2015rva,Dror:2018pdh,Agrawal:2018vin,LowEnergyFrontierReview2010, Ahmed:2020fhc,GravitationalProductionOfDP2021,Co_2019,Co_2021}. 
Such a DP will generically have a kinetic mixing with the SM photon because this term is allowed by the symmetries of the theory (so long as the DP does not have a non-Abelian gauge symmetry). This kinetic mixing allows dark photon dark matter (DPDM) to be looked for in existing~\cite{DPLimitsReview2021,HuntForDarkPhoton2020} and forthcoming experiments~\cite{Antypas:2022asj}.  

In this Letter, we propose a promising new direct detection technique using one-quantum transitions of one or more trapped electrons that are initially cooled to their cyclotron ground state.  We demonstrate the viability of this technique with a proof-of-principle measurement that sets a limit 75 times better than previous constraints.  This new limit is only for a narrow mass range because of limitations of an apparatus designed for making the most accurate measurements of the electron and positron magnetic moments~\cite{atomsNewMeasurement2019}---to test the standard model's most precise predictions~\cite{HarvardMagneticMoment2008,HarvardMagneticMoment2011,Nio2018TenthOrder,CsAlphaScience2018,RbAlpha2020Nature,Alpha2006fixed,atomsTheoryReview2019,DehmeltMagneticMoment,EfficientPositronAccumulation}.  With an apparatus designed for DPDM detection, including efficient scanning of the resonant frequency, the mass range could be greatly extended.

The relevant properties of the DP are captured by the Lagrangian (in natural units)~\cite{DPTheory1986}
\begin{align}
    \mathcal{L}\supset -\frac{1}{4}F'_{\mu\nu}F'^{\mu\nu} + \frac{\epsilon}{2} F^{\mu\nu}F'_{\mu\nu}+\frac{1}{2}m_{A'}^2A'_\mu A'^\mu.
\end{align}
Here $A'_\mu$ is the DP vector, $F'$ and $F$ are the DP and SM photon field strengths respectively, $\epsilon$ is the kinetic mixing parameter, and $m_{A'}$ is the mass of the DP. The DPDM manifests as dark electric and magnetic fields oscillating at a frequency set by the DP mass $\omega_{A'}=m_{A'}c^2/\hbar$, where $c$ is the speed of light and $\hbar$ is the reduced Planck constant. In the presence of a kinetic mixing with the SM photon, these dark fields cause effective ($\epsilon$-suppressed) SM electromagnetic fields. These can be detected by devices sensitive to tiny electric or magnetic fields at the frequency $\omega_{A'}$. 

A plethora of complementary experiments have been designed with sensitivities to different DM masses.
The frequency range we focus on, 20~to~200~GHz (i.e.,\ 0.1 to 1~meV) is particularly challenging experimentally, yet well-motivated theoretically by the minimal DPDM model with purely gravitational production \cite{Graham:2015rva}.  This range is too high for extremely high-$Q$ resonators (e.g.,~as used by ADMX  \cite{ADMXSideCar2018,ADMXTechnicalDesignReview2021}, CAPP \cite{CAPP2020_7ueV,CAPP2020_13ueV,CAPP2021_10ueV}, and HAYSTAC \cite{HAYSTAC2018_24ueV,HAYSTAC2021_17ueV}).  At the same time, the corresponding photons are below the energy threshold for existing single photon detection experiments such as those in Refs.~\cite{Chiles:2021gxk,Hochberg:2021yud,SinglePhotonReview2011}. Alternate experiments involving dish antennae or metal plates have been proposed or are underway around our frequency range \cite{DOSUE2022_100ueV, DarkPhoton_Tomita2020_115ueV, DarkPhoton_Knirck_2018_0.8meV, BREAD:2021tpx}.  
The use of trapped ion crystals was proposed for the MHz frequency range \cite{gilmore2021quantum}.

\begin{figure}[tb!]
    \centering
    \includegraphics[width=\the\columnwidth]{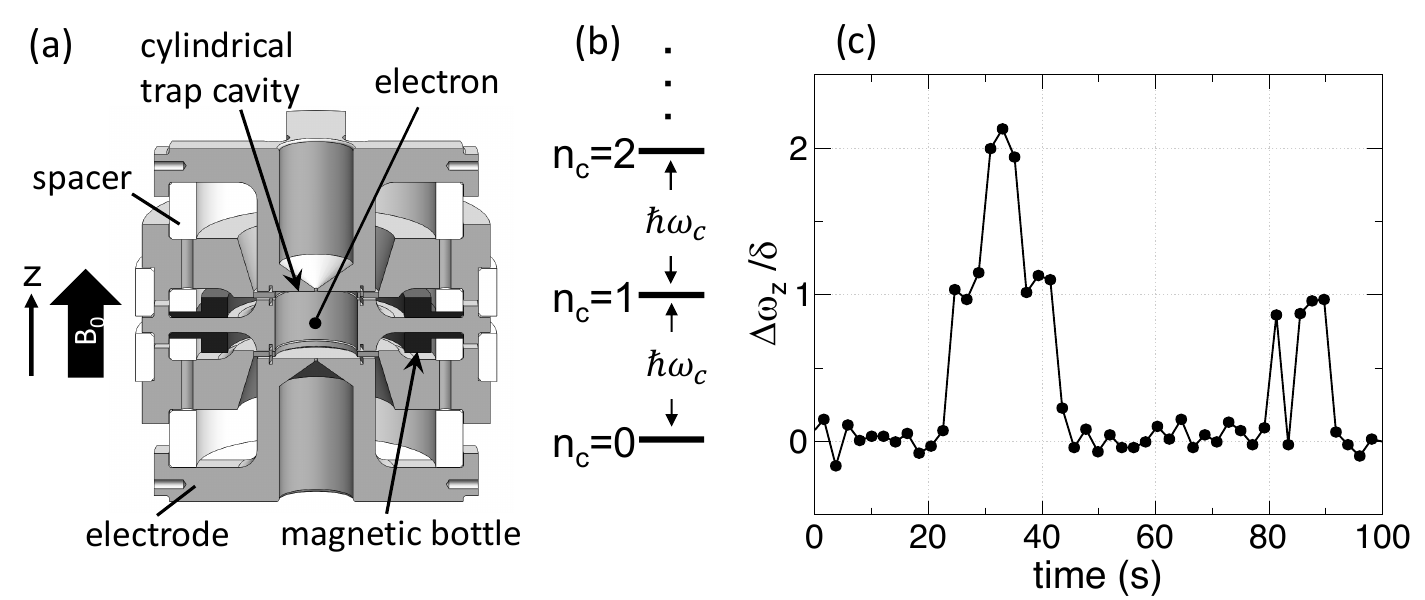}
    \caption{(a) Single isolated electron in a cylindrical Penning trap. (b) Quantum cyclotron energy levels of the trapped electron. (c) Resolution of cyclotron states $n_c$ by measuring axial frequency shift. A resonant external cyclotron excitation drive is applied to cause stimulated transitions of $n_c$.}
    \label{fig:TrapAndQuantumStates}
\end{figure}

The new DPDM detector proposed and demonstrated here is one (or more) electron in a Penning trap [Fig.~\ref{fig:TrapAndQuantumStates}(a)]---a ``one-electron quantum cyclotron'' \cite{QuantumCyclotron}.  The trapped electron is a high-$Q$ resonator with a 20--200~GHz resonant frequency determined by the applied magnetic field of the trap.   The DPDM wave would drive the electron to jump from the cyclotron ground state to the first excited state [Fig.~\ref{fig:TrapAndQuantumStates}(b)] if the cyclotron level spacing corresponds to the DP's frequency.  The cyclotron quantum state is monitored in real time [Fig.~\ref{fig:TrapAndQuantumStates}(c)] to search for excitations.  To determine the cyclotron excitation rate, we compute the SM electric field induced by the DPDM inside the microwave cavity formed by the electrodes of the Penning trap. Similar techniques were employed in the context of ion traps to search for millicharge dark matter in Ref.~\cite{Budker:2021quh}.

Two motions of the trapped electron (mass $m_e$ and charge $-e$) are key.  The quantized cyclotron oscillation (in a plane perpendicular to a strong magnetic field $B_0 \hat{z}$) is potentially excited by the DPDM-generated photon field in the $xy$ plane. For $B_0 = 5.3$~T, a photon resonant at the cyclotron frequency, $\omega_c/(2\pi) = eB_0/(2\pi m_e)=148$~GHz could increase the cyclotron energy by one quantum, $\hbar \omega_c$ [Fig.~\ref{fig:TrapAndQuantumStates}(b)].  The frequency of the electron's classical axial oscillation, along the magnetic field, is used to detect cyclotron excitations~\cite{DehmeltMagneticBottle}.
The axial oscillation frequency is set at $\omega_z/(2\pi)=114$~MHz for $n_c=0$ by the static potentials applied to the trap electrodes. 

\begin{figure}[tb!]
    \centering
    \includegraphics[width=0.9\columnwidth]{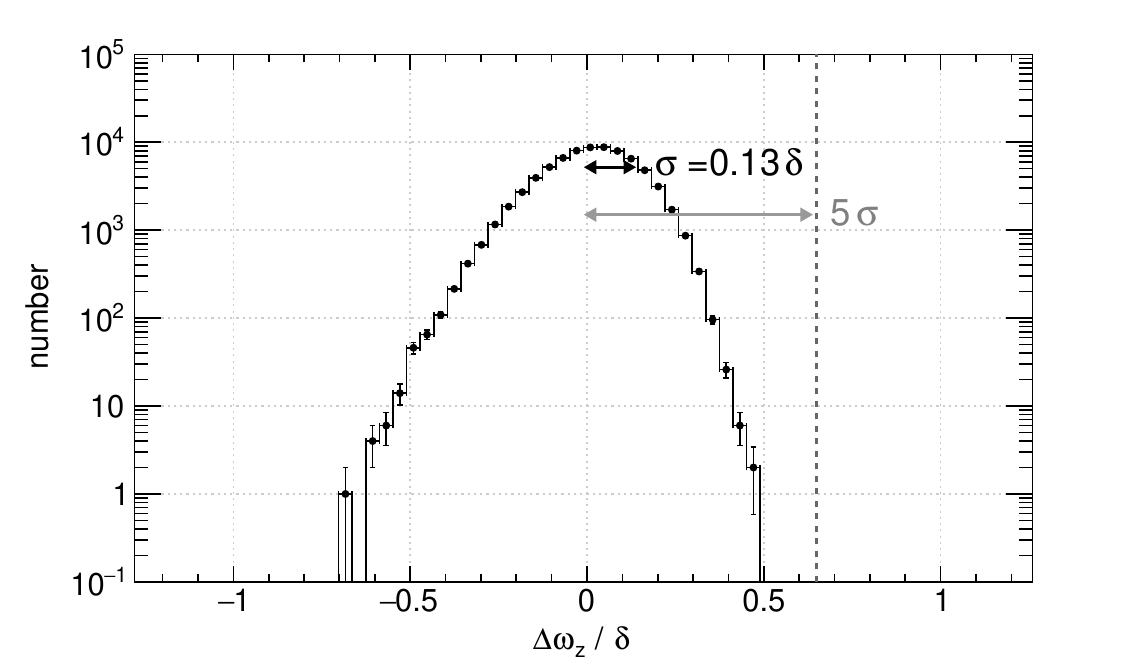}
    \caption{Distribution of measured axial frequency fluctuations $\Delta\omega_z/\delta$ showing the standard deviation $\sigma$ and the chosen threshold at $5\sigma$.}
   \label{fig:DistributionOfMeasuredNc}
\end{figure}

A quantum nondemolition coupling of the two motions makes it possible to detect one-quantum cyclotron excitations without causing a change to the cyclotron quantum number~\cite{FanBackActionPRL2021}. The monitored axial frequency shifts in proportion to the cyclotron quantum number $n_c$ by $\Delta \omega_z = n_c \delta$ [Fig.~\ref{fig:TrapAndQuantumStates}(c)], due to a magnetic bottle gradient that adds $B_2z^2 \hat{z}$ to the magnetic field.
The potential along $z$ now includes a quadratic magnetic potential $V(z)={m_e\omega_z^2z^2}/{2}+\hbar eB_2n_cz^2/m_e$~\cite{DehmeltMagneticBottle} given by the product of the cyclotron orbital magnetic moment and $B_2z^2$.
A nickel ring encircling the trap generates $B_2=300$~T/m$^2$, making $\delta/(2\pi) \equiv \hbar eB_2/(2\pi m_e^2\omega_z) = 1.3$~Hz.
The axial frequency is read out by keeping the axial oscillation amplitude at large $z_\text{max}=60~\mu\mathrm{m}$ by feeding back an electrical signal induced by the oscillation itself~\cite{SelfExcitedOscillator}.
The axial shift $\delta/(2\pi)$ from a one-quantum cyclotron excitation is 8 times larger than the $\sigma/(2\pi) = 0.16~\mathrm{Hz}$ standard deviation for fluctuations, predominantly from the axial amplitude fluctuation coupled to the trap potential's anharmonicity\cite{SelfExcitedOscillator}. 
The axial detection is continuously on, but the coupling is too weak to cause the quantum Zeno effect and suppress the cyclotron transitions\cite{QuantumZenoEffectWineland,ThesisPeil}.
The distribution of measured axial frequencies for 2~s averaging time is displayed in Fig.~\ref{fig:DistributionOfMeasuredNc}.

Because of the $B_2$ coupling, the cyclotron resonance frequency is also broadened to $Q=\omega_c/\Delta\omega_c=B/(B_2z_\text{max}^2)\approx10^7$ from the unbroadened $Q=10^9$  \cite{HarvardMagneticMoment2008}.
This does not worsen the smallest $\epsilon$ we can prove, and slightly broadens the  DPDM sensitivity bandwidth beyond the intrinsic $Q=10^6$ bandwidth of the dark matter \cite{Bovy_2012_DMVelocity,Schonrich_DMVelocity,Golubov_DMVelocity}.

The electron is suspended at the center of a trap~[Fig.~\ref{fig:TrapAndQuantumStates}(a)] that a dilution refrigerator keeps at a temperature of $T=50$~mK.  The electron cyclotron motion cools via synchrotron radiation and is not excited by blackbody photons.  
In thermal equilibrium, the average quantum number is \cite{QuantumCyclotron}
\begin{equation}
    \bar{n}_c = \left[\exp\left(\frac{\hbar\omega_c}{k_\mathrm{B}T}\right)-1\right]^{-1}= 1.9\times10^{-62}\approx 0,
\end{equation}
where $k_B$ is the Boltzmann constant.
The electron is thus essentially always in its quantum cyclotron ground state $n_c=0$, with no background excitations from blackbody photons estimated to take place for many years \cite{QuantumCyclotron}.

Photons dynamically induced by the DPDM field could produce cyclotron transitions from $n_c=0\rightarrow n_c=1$.
For radiation broader than the linewidth, the transition rate is~\cite{loudon2000quantum,NoiseDrivenTransition}
\begin{equation}
    \Gamma= \int \frac{\pi e^2}{2 m_e \hbar\omega_c} S_E\left(\omega\right)\chi(\omega,\omega_c)d\omega,
    \label{eq:GammaDPDM}
\end{equation}
where $\chi(\omega,\omega_c)=\tfrac{1}{\sqrt{2\pi}\Delta\omega_c}\exp\left[-\tfrac{1}{2}\left(\tfrac{\omega-\omega_c}{\Delta\omega_c}\right)^2\right]$ is the normalized cyclotron line shape with linewidth $\Delta\omega_c$, and 
$S_E(\omega)\,d\omega$ is the power in the interval $\{\omega,\omega+d\omega\}$
for the component of the DM-induced electric field in the $xy$ plane.  
For DPDM with spread $\Delta\omega_{A'} \approx 10^{-6}\omega_{A'}$~\cite{DMDispersion2000}, and for a cylindrical cavity, $S_E(\omega)$ can be approximated as a boxcar window function with value 
\begin{align}
S_E(\omega)=\kappa^2\times\epsilon^2\frac{\rho_{\rm DM}c^2}{\varepsilon_0\Delta\omega_{A'}} \langle \sin^2\theta \rangle
\label{eq:SE_DPDM}
\end{align} 
in the interval $\{\omega_{A'},\omega_{A'}+\Delta\omega_{A'}\}$ and zero outside, where $\rho_\mathrm{DM}c^2=0.3$~GeV/cm$^3$ is the local DM density~\cite{Planck2018-1} and $\varepsilon_0$ is the vacuum permittivity. We assume that the angle between the DPDM electric field and the $z$ axis, $\theta$, changes randomly
and is adequately sampled in observation times $T_{\rm obs}\gg {1}/{\Delta\omega_{A'}}\approx 10^{-6}~\mathrm{s}\times \left(\frac{2\pi\times 148 ~\textrm{GHz}}{\omega_{A'}}\right)$. 
Thus the angular average that captures the component along the $xy$ plane evaluates to
$\langle \sin^2\theta \rangle =2/3$.  
A fixed DPDM polarization~\cite{DPLimitsReview2021} would essentially not change the result given that our apparatus, off the Earth's rotation axis, changes orientation as the Earth rotates during the $T_{\rm obs}\gg $ 1 day observation time for this experiment~\cite{forthcoming}.
We calculate the effect of fixed DPDM polarization on our projected future sensitivity in Ref. \cite{forthcoming}.

Finally, $\kappa$ is the enhancement of the DPDM-induced electric field at the position of the electron by the trap's microwave structure~\cite{DMRadioPRD_2015}:
\begin{equation}
    \kappa=\left|\sum_n\frac{\omega^2}{\omega^2-\omega_n^2(1-\tfrac{2i}{Q_n})}\frac{\int dV\vec{E}_n^*(\bf{r})\cdot \hat{\mathbf{x}} }{\int dV|\vec{E}_n(\bf{r})|^2}\vec{E}_{n}(\bm{0})\cdot \hat{\mathbf{x}} \right|.
    \label{eq:KappaExplicitExpression}
\end{equation}
Without loss of generality, $\mathbf{x}$ is taken to be the DPDM polarization direction; 
$n$ runs over all resonant modes; and $\omega_n$, $Q_n$, and $\vec{E}_n(\bf{r})$ are the resonant frequency, quality factor, and electric field of the mode at position $\bf{r}$ respectively.
The last term $\vec{E}_{n}(\bm{0})\cdot \hat{\mathbf{x}}$ captures the transverse electric field at the center that  drives electron cyclotron transition.

\begin{figure}[]
\centering
  \includegraphics[width=\columnwidth]{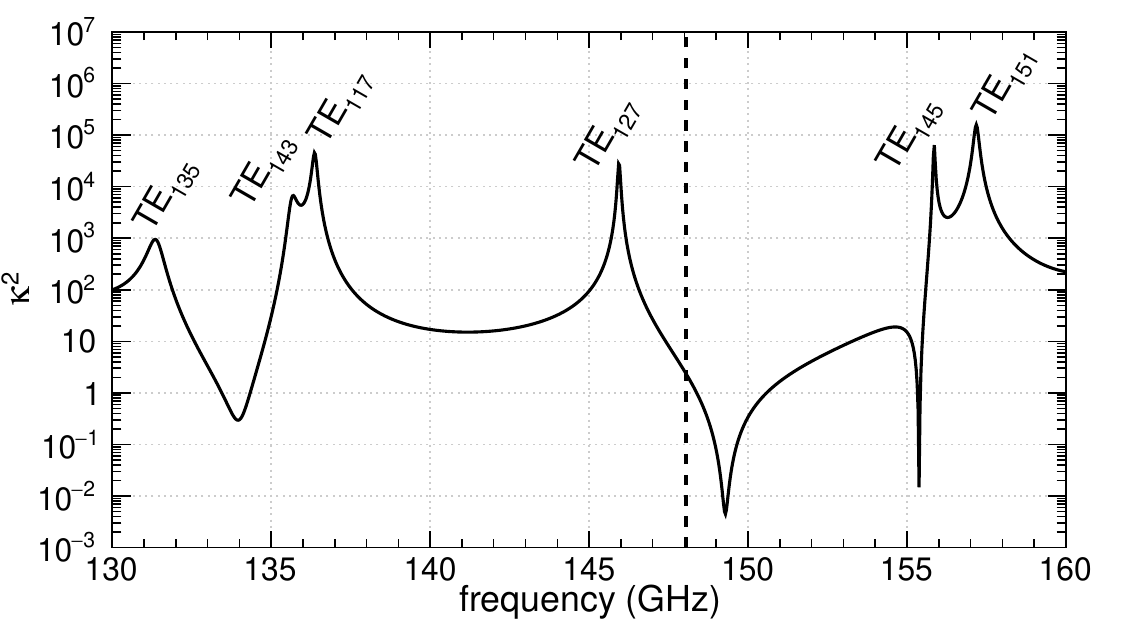}
 \caption{Calculated $\kappa^2$ for the measured microwave resonant frequencies and $Q$ factors. For this demonstration search (at 148.04786~GHz), $\kappa^2 = 2.37$ (dashed line).}
  \label{fig:kappaplot}
\end{figure}

Figure~\ref{fig:kappaplot} shows the calculated frequency spectrum for $\kappa^2$ using measured resonant frequencies and $Q$ factors.  The sharp peaks are from cavity modes that couple strongly to the cyclotron motion of an electron suspended at the cavity center. The microwave cavity resonances below 170~GHz for the cylindrical Penning trap \cite{CylindricalPenningTrap,CylindricalPenningTrapDemonstrated}  (with radius $\rho_0=4.527$~mm and height $2z_0=7.790$~mm) have all been carefully mapped using parametrically pumped electrons~\cite{SynchronizedElectronsPRL,SynchronizedElectronsPRA}.  The measured frequencies agree with an ideal cylindrical model to within a few percentages.

The high $\kappa$ values at cavity mode resonances are unfortunately not compatible with the existing $2 \, \text{s}$ averaging time needed to resolve the one-quantum cyclotron transitions in our apparatus. 
For this averaging time, the  magnetic field must be chosen to keep the electron cyclotron frequency far from resonance with all cavity modes that couple to the centered electron.
This inhibits the spontaneous emission of synchrotron radiation to lengthen the lifetime of the excited cyclotron state $\tau_c$ \cite{InhibitionLetter}. For this demonstration, $\tau_c$ was set at the longest value, $\tau_c = 7.2$~s, a factor of 80 longer than its free space value \cite{InhibitionLetter,HarvardMagneticMoment2011}, to avoid missing a signal.
The photons induced by the DPDM, being away from resonance, thus cannot build up in a cavity radiation mode.  Fortunately, our calculation shows that $\kappa^2$ remains usefully large, with $\kappa^2=2.37$ pertaining to our demonstration at 148~GHz. The cylindrical symmetry of the conducting cavity boundary causes an enhancement in the DM-induced electric field at the trap center (akin to the ``focusing'' effect found in the dish antenna proposals \cite{DishAntennaProposal-1,DishAntennaProposal-2}). A future optimization by carefully choosing $\tau_c$ and by using a larger spherical trap cavity 
can result in a 25-fold increase in $\kappa$.

\begin{figure}[]
    \centering
    \includegraphics[width=\the\columnwidth]{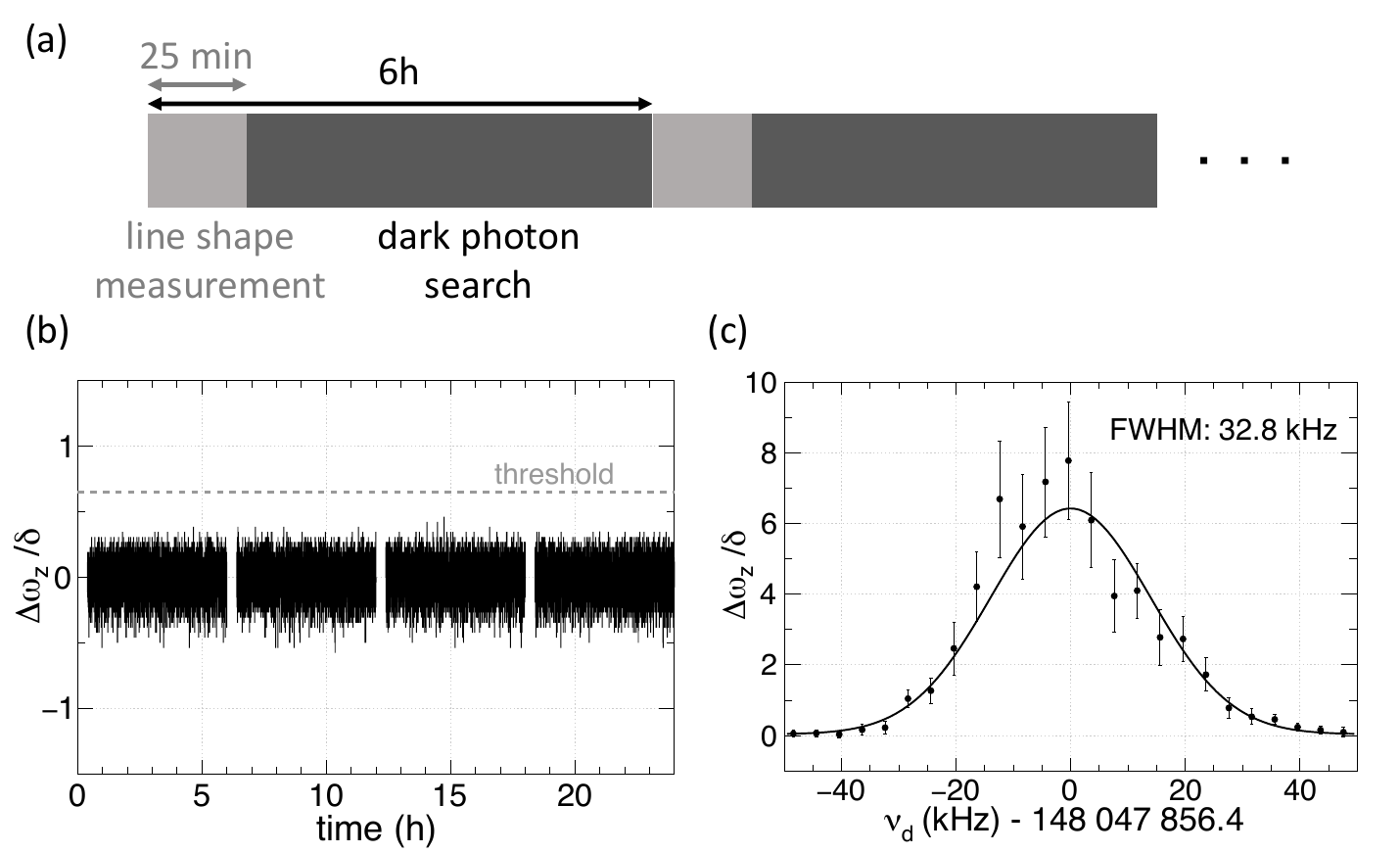}
    \caption{(a) The DPDM measurement cycle. (b) Monitored axial frequency shift shows no DPDM-induced cyclotron excitations during 24 hours of the 7.4 day search. (c) Measured cyclotron line shape $\chi(\omega,\omega_c)$ with the self-excitation on.}
    \label{fig:MeasurementCycle}
\end{figure}

The new search for 148~GHz DPDM [Fig.~\ref{fig:MeasurementCycle}(a)] is for $n_c=0$ to $n_c\ge 1$ cyclotron excitations over $T_\mathrm{obs} = 7.4$~days (Table~\ref{tab:datasets}). Shifts of the electron's axial frequency are averaged over $t_\mathrm{ave}=2$~second intervals and recorded as a function of time.  The trapping potential is slowly adjusted to eliminate slow drifts of the axial frequency.  Figure~\ref{fig:MeasurementCycle}(b) shows $\Delta \omega_z/\delta$ for 24 hours of the 7.4 day search. A cyclotron excitation to the first excited state would produce $\Delta \omega_z/\delta = 1$.  Any $\Delta \omega_z$ larger than a $5\sigma$ threshold (i.e.\ $\Delta \omega_z/\delta = 5\sigma/\delta \ge 0.65$) would be interpreted as being potentially caused by DPDM. No such excitation is detected during the 7.4 days.

The search was suspended for 25~min of calibration every 6 hours, indicated by the breaks in Fig.~\ref{fig:MeasurementCycle}(b).
The one-quantum response is confirmed using microwave photons sent into the cavity\cite{HarvardMagneticMoment2011}.
The detector bandwidth of 33~kHz is also deduced by measuring the shift $\Delta \omega_z /\delta$ as a microwave drive is swept through resonance with the 148~GHz cyclotron frequency [Fig.~\ref{fig:MeasurementCycle}(c)]. The width is broadened by the large self-excited axial oscillation in the magnetic gradient described above. Cyclotron frequency shifts are negligible given the extremely low magnetic field drift rate of $\Delta B/B=10^{-10}$ per hour that is realized using a carefully shimmed self-shielding solenoid \cite{SelfShieldingSolenoid,Helium3NMR2019,atomsNewMeasurement2019}.  This 33 kHz detector bandwidth slightly broadens the sensitivity bandwidth beyond the ${\sim}100 \, \text{kHz}$ bandwidth expected of the dark matter.

\begin{table}[]
\begin{tabular}{c|c|c}
\hline
Run No. & Time (date.hour:minute) &Observation length (s)  \\
\hline
\hline
1     & 11.12:46~ -- ~13.13:15     &148058   \\
2     & 14.18:26~ -- ~15.11:33     &58162   \\
3     & 15.11:50~ -- ~17.17:22     &179698   \\
4     & 17.18:38~ -- ~18.18:40     &80640   \\
5     & 19.12:15~ -- ~21.15:43     &172312   \\
\hline\hline
Total     & $\cdots$     &638870   \\
\end{tabular}
\caption{Datasets for DPDM search in March 2022. Each run consists of the repeated measurement cycle in Fig.~\ref{fig:MeasurementCycle}.}
\label{tab:datasets}
\end{table}

The lowest cyclotron excited state decays to the ground state by the spontaneous emission of synchrotron radiation photons. The decay time for each excitation is a random selection from an exponential distribution with an average lifetime of $\tau_c =7.2$~s. The choice of a detection threshold at $5\sigma=0.65\delta$ means that an excitation that decays in less than $0.65\times t_\mathrm{ave}=1.3$~s will be missed, giving a detection efficiency of
\begin{equation}
\zeta = \int_{1.3\mathrm{s}}^\infty\frac{1}{\tau_c}\exp\left(-\frac{t}{\tau_c}\right)dt=83\%.
\end{equation}
Correcting for the observed fluctuation spectrum in  Fig.~\ref{fig:DistributionOfMeasuredNc} affects the result by only 1\%.

\begin{figure*}[t]
\centering
    \begin{minipage}{\columnwidth}
        \centering
  \includegraphics[width=\columnwidth]{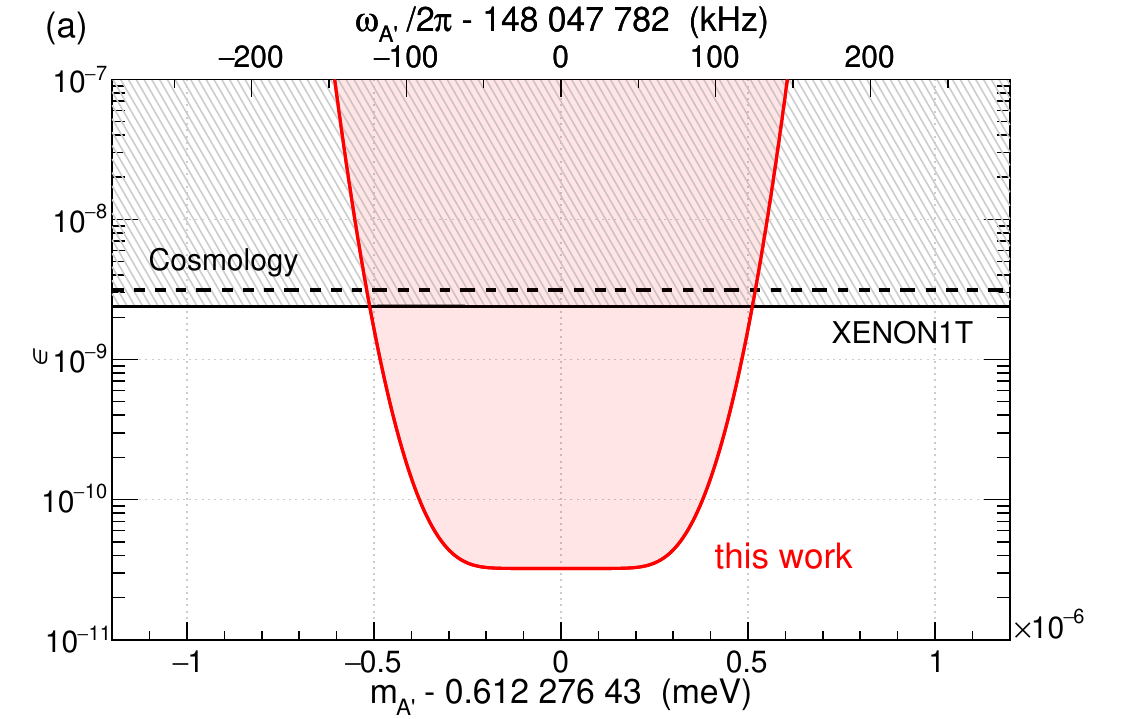}
\end{minipage}\hfill
    \begin{minipage}{\columnwidth}
        \centering
\includegraphics[width=\columnwidth]{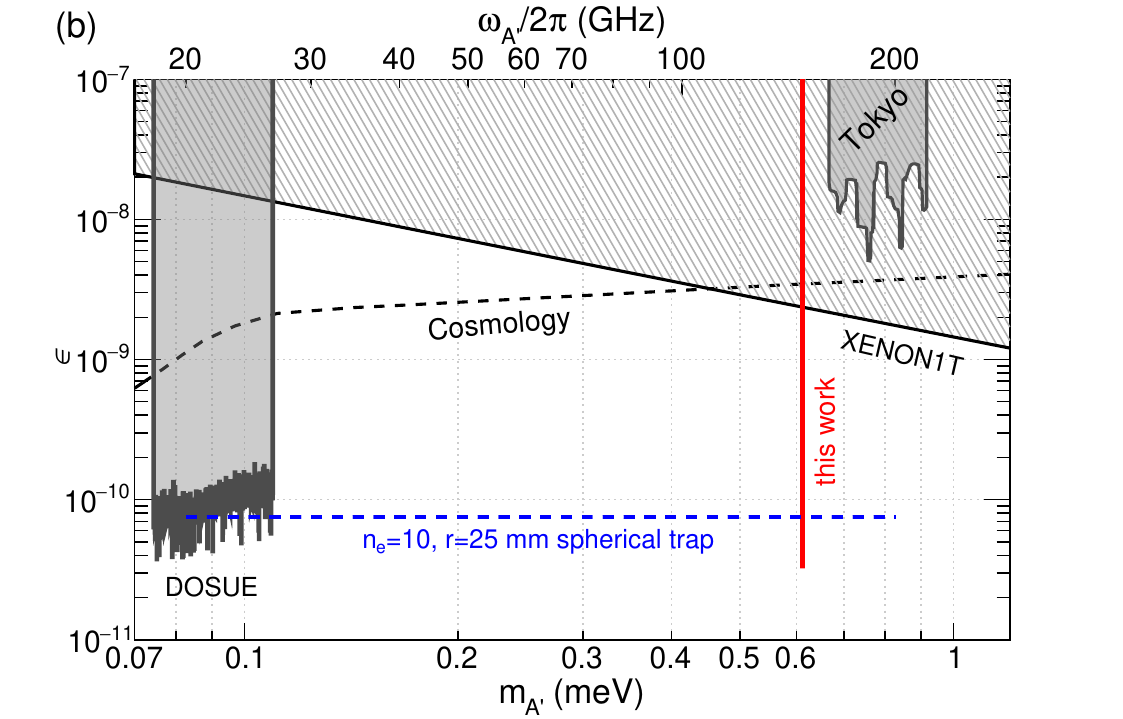}\end{minipage}\hfill
  \caption{(a) The new experimental limit on DPDM with 90~\% confidence level.  In a narrow frequency band, the exclusion limit from this work is about 75 times better than the existing constraints from XENON1T (hatched)~\cite{DMDetectorAsDPHelioscope,Xenon1TResultDarkPhoton,Xenon1TPRL2019} and DM cosmology~(dashed)~\cite{WISPyDarkMatter2012}. 
  (b) Possible sensitivity with a dedicated apparatus (blue) is compared to the  limits from DOSUE~\cite{DOSUE2022_100ueV} and Tokyo~\cite{DarkPhoton_Knirck_2018_0.8meV}.}
    \label{fig:LimitOnDPDM148GHz}
\end{figure*}

The conversion to $\Gamma$ is now straightforward. Using the standard estimate of the upper limit of the null measurement \cite{LeoParticlePhysicsTextBook}, the upper limit on the DPDM excitation rate with $CL=90$\% confidence level is
\begin{equation}
    \Gamma < -\frac{1}{\zeta T_\mathrm{obs}}\log\left(1-CL\right)=4.33\times10^{-6}~\mathrm{s}^{-1},
\end{equation}
which is also the upper limit of the one-electron detector's dark count rate.
Using measured values and Eqs.~\eqref{eq:GammaDPDM} and \eqref{eq:SE_DPDM}, our limit on the kinetic mixing parameter is
\begin{equation}
\epsilon < 3.2\times10^{-11}  
\end{equation}
at $m_{A'} =0.612\,276$~meV, which improves on previous limits by a factor of 75. 
The corresponding microwave electric field detected, specified by $\sqrt{2\pi S_E(\omega)}$ is given by $2.5~\mathrm{pV/(cm}\sqrt{\text{Hz}})$, with the measurement  bandwidth, 0.45~nV/cm. 
The new $\epsilon$ limit is shown in Fig.~\ref{fig:LimitOnDPDM148GHz}(a), with the limit from the XENON1T (black hatched) \cite{DMDetectorAsDPHelioscope,Xenon1TResultDarkPhoton,Xenon1TPRL2019} and the limit from DM cosmology (dashed) \cite{WISPyDarkMatter2012}.

Only a narrow DPDM mass range is accessed in this initial demonstration due to limitations of the apparatus, which was designed to make the magnetic field exceptionally stable rather than readily swept.
Searching 20--200~GHz ($\sim$0.1~meV to 1~meV) seems feasible in an apparatus that is designed for dark matter searches.  Affordably sweeping the magnetic field over such a broad range requires cooling with a refrigerator rather than cryogenic liquids.   
For DM with $Q=10^6$, making $t_m=15$~s measurements spaced by $10^{-6}$ relative frequency steps would cover the mentioned range in about a year. 
The DPDM sensitivity established above is approximately 
\begin{align} \label{eq:futurelim}
        \epsilon\approx8\times10^{-11}\frac{\omega_{A'}}{2\pi\times150~\textrm{GHz}}\frac{40}{\kappa}\left(\frac{10}{n_{\rm e}}\right)^\frac{1}{2} \left(\frac{15~\textrm{s}}{t_{\rm m}}\right)^\frac{1}{2},
\end{align}
where $n_e$ is the number of electrons used to sense DM and $t_m$ is the measurement time. A shorter measurement time (15~s rather than 7.4~days) would decrease the sensitivity $\epsilon$ by a factor of 200. It seems feasible to largely recapture this factor by using $n_e=10$ electrons, increasing $\kappa$ to ${\sim} 40$ by using a spherical geometry to get more focusing, and increasing the radius of the sphere to $r=25~\textrm{mm}$~\cite{forthcoming}. 
Resulting reductions in the induced axial oscillation signal needed to observe one-quantum cyclotron jumps would be compensated for by greatly increasing the size of the magnetic bottle gradient that couples the cyclotron and axial motion. The blue dashed line in Fig.~\ref{fig:LimitOnDPDM148GHz}(b) is an estimate of what may be possible, assuming that the trap cavity can be tuned during the sweep to avoid cavity mode resonances. For a spherical cavity, $\kappa \propto \frac{\omega_{A'}r}{c} $ which cancels the $\omega_{A'}$ in Eq.~\eqref{eq:futurelim}, leading to a flat sensitivity curve. A more detailed optimization is clearly warranted \cite{forthcoming}.

In conclusion, we have proposed and demonstrated the possibility of using one-electron quantum cyclotrons within a microwave trap cavity to search for DPDM. A big advantage is that detection is essentially free of background excitations, making the detection sensitivity of the transition rate scale with observation time as $T_\mathrm{obs}^{-1}$ rather than $T_\mathrm{obs}^{-\frac{1}{2}}$.  The obtained limit is the most sensitive ever obtained in the challenging meV range and all required parameters for the DPDM search are measured \textit{in situ}. The narrow frequency range realized in this first demonstration could be greatly extended in an apparatus designed and optimized for dark matter detection.
Furthermore, the trapped electron is naturally compatible with a large magnetic field, and application to an axion search is promising with a dedicated antenna \cite{forthcoming}.  
This proposal and demonstration thus open a new direction for DPDM searches.

\begin{acknowledgments}

This work was supported by the U.S. DOE, Office of Science, National QIS Research Centers, Superconducting Quantum Materials and Systems Center (SQMS) under Contract No.\ DE-AC02-07CH11359. Additional support was provided by NSF Grants No.~PHY-1903756, No.~PHY-2110565, and No.~PHY-2014215; by the John Templeton Foundation Grants No.~61906 and No.~61039; by the Simons Investigator Award No.~824870; by the DOE HEP QuantISED Award No.~100495; by the Gordon and Betty Moore Foundation Grant No.~GBMF7946; and by the Masason Foundation. S.W.~was supported in part by the Clark Fellowship. Y.X.~was supported in part by the Vincent and Lily Woo Fellowship.
\end{acknowledgments}

\bibliographystyle{prsty_gg}
\bibliography{PenningTrapExperimentRefs}

\end{document}